# Claudius Ptolemy (ca. AD 100 – ca. 170) and Giambattista Della Porta (ca. 1535 –1615): Two Contrasting Conceptions of Optics

Yaakov Zik and Giora Hon
*Department of philosophy, University of Haifa, Israel*

## Abstract

We address the phenomenon of reflection in concave spherical mirror in two contrasting approaches to optics. In his *Optics* (ca.165) Ptolemy applied the cathetus principle as a regulative means for explaining qualitatively effects related to visual perception in concave spherical mirror. By contrast, Della Porta's study of reflection in concave spherical mirror in Bk. 17, Ch. 4 of his *Magia naturalis* (1589) and *De refractione* (1593), was based on the assumption that there is a reciprocal relation between reflection in concave spherical mirror and refraction in glass sphere. We juxtapose these two studies and draw several philosophical lessons from the comparison between these two practices with a view to throwing into relief the fundamental differences in their respective conceptions of optics.

**Claudius Ptolemy (ca. AD 100 – ca. 170) and Giambattista Della Porta (ca. 1535 –1615): Two Contrasting Conceptions of Optics**

## 1. Introduction

This paper is a comparative study. We juxtapose the optical work of Ptolemy with that of Della Porta. Specifically, we are concerned with the phenomenon of reflection in concave spherical mirrors. Our goal is philosophical: to compare two distinct cases of scientific practice within the domain of optics which address the same phenomenon. The comparison, we claim, throws into relief two contrasting conceptions of optics. In the case study we analyze, Ptolemy inquired into the image location in concave spherical mirror and its dependence on the position of the eye in relation to the perceived object, while Della Porta inquired into the location where parallel light rays striking the concave spherical mirror would unite and congregate most densely. These then are two distinct inquiries into the same phenomenon. In other words, we take two snapshots of scientific practices and compare them. The juxtaposition is intended to highlight the fundamental difference in the way optics was conceived. We believe revealing lessons can be drawn from this contrast. We do not seek to situate these two episodes in the history of optics; indeed, we are not concerned with a historical development which would link the work of Ptolemy with that of Della Porta across a millennium and a half. And we do not judge whether the results obtained by these two illustrious scholars are correct. This is not our brief. We argue that what differentiates the two practices is that Ptolemy’s methodology is essentially “subjective” (based on what is seen in the mirror) whereas that of Della Porta is essentially “objective” (based on how light interacts physically with the reflecting surface).

We ask, What was Ptolemy’s practice, and what was the methodology which Della Porta applied? Once we possess answers to these questions we will be able to characterize the similarities and differences between these two scientific practices in the domain of optics. A short account of Greek optics serves as a background for an exposition of Ptolemy’s study of concave spherical mirrors; we then move on to Della Porta and his study of reflection in such mirrors. In the conclusion we execute the comparison and draw lessons relevant to the respective conception of optics of Ptolemy and Della Porta.

## 2. Historical background: Greek optics and Euclid's scheme of concave spherical mirrors

Greek optics sought to explain, inter alia, visual perception and its accompanied optical deceptions as perceived by the eye. All Greek visual theories presupposed that vision cannot occur without a contact mediated by the visual ray suspended between the eye and the visible object.[1]

The Greeks developed two fundamental processes of vision:[2]

[1] For background discussion, see Smith, 1996; 1999; 2015.

[2] On the visual ray and ancient theories of vision, see Smith, 1999, pp. 23–34. On the foundation of classical geometrical optics, see Smith, 1981.

1. A theory which presupposes the passage of something from the object into the eye (Aristotle and the Atomists) ─ Intromission;
2. A theory which presupposes the passage of something from the eye to the object (Pythagoreans, Euclid, and Ptolemy) ─ Extramission.

Euclid's optical theory is based on the following definitions and postulates (Cohen and Drabkin, 1966, pp. 257─258; Burton, 1945, p. 357; Smith, 1999, pp. 51–58).

1. The rectilinear visual rays proceeding from the eye diverge indefinitely.
2. The figure contained by a set of visual rays is a cone of which the vertex is at the eye and its base at the surface of the object seen.[3]
3. Objects within the visual cone are seen and objects outside of the visual cone are not seen.
4. Objects seen under greater angle will appear greater, objects seen under smaller angle will appear smaller, and objects seen within equal angles appear to be of the same size.
5. Objects seen under higher visual ray will appear higher, and objects seen under lower visual ray will appear lower.
6. Objects seen within the visual cone on the right appear on the right, while objects seen within the visual cone on the left appear on the left.
7. Objects seen under larger angles will appear larger and more clearly.

The representation of the visual rays as straight lines diverging from a point in the eye reduces optics into purely geometrical propositions which had been already discussed in the *Elements*. In effect, Euclid's *Optics* offered a scheme for following the correspondences between an object and its image in the eye in terms of the path of the rays within the visual cone according to the following rules,

1. Visual rays link geometrically the observer's eye with the object in space.
2. Geometrical entities (lines, arcs, triangles) are used to define the relation between the observer's eye and the object.
3. Arithmetical operations convert these relations to physical magnitudes such as distance, height, and apparent angle.

The visual-ray model embodies the necessary geometrical elements for constructing a satisfactory theory of visual perception.[4]

In yet another optical text, the *Catoptrics*, Euclid sought to explain aspects of visual appearances in mirrors. Euclid did not consider color a factor; color does not affect the visual ray. In the following propositions Euclid setup the scheme for determining the location of images reflected by mirrors of various shapes.[5]

---

[3] In effect, the eye radiates visual flux in the shape of a cone, whose vertex lies within the eye to define the center of sight, and whose base defines the visual field. This visual cone can be resolved into a bundle of discrete rays, each one projected outward from the vertex.

[4] This model however is problematic in the sense that it ignores color perception and it does not include psychological factors (e.g., the Moon illusion).

[5] For Propositions 16–18 see Euclid, 1537, pp. 79–80; Smith, 1999, pp. 96–97.

Euclid’s *Catoptrics*, Proposition 16 (Fig. 1):
In a plane mirror the image (E) is seen behind the mirror at the intersection of the cathetus (perpendicular) (OE) drawn from the object to the mirror and the extension of the incident ray (BDE).

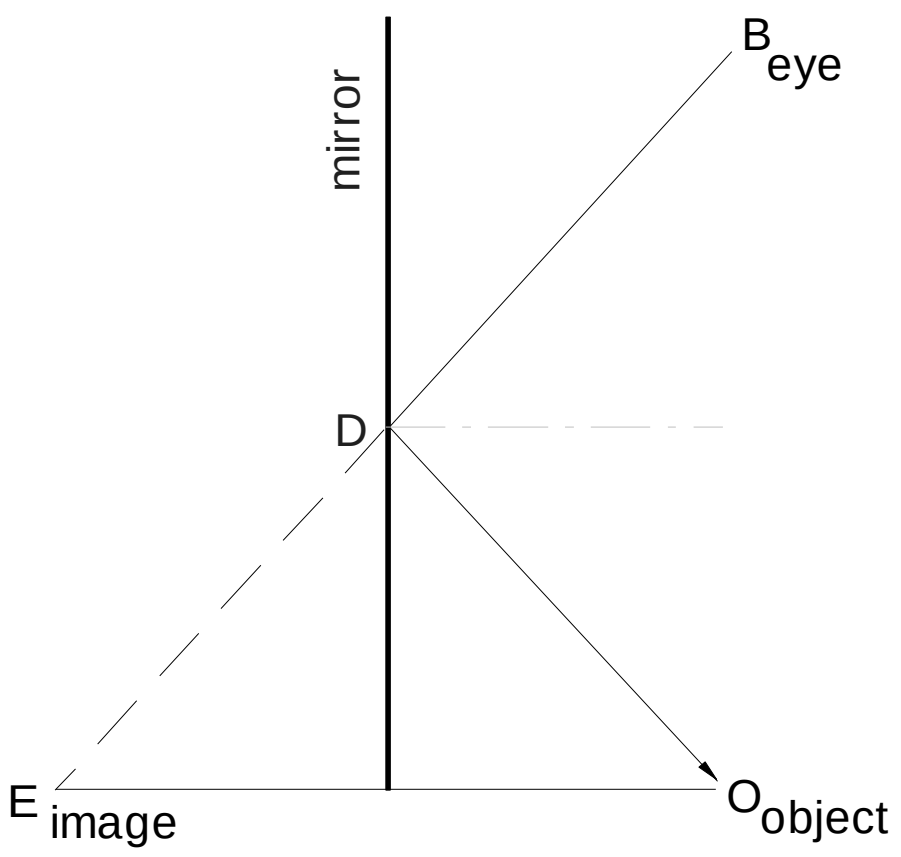

Figure 1

Euclid’s *Catoptrics*, Proposition 17 (Fig. 2)
In a convex spherical mirror the image (E) is seen behind the mirror at the intersection of the cathetus (perpendicular) (AZ) drawn from the object (A) to the center of the sphere (Z) and the extension of the incident ray (BDE).

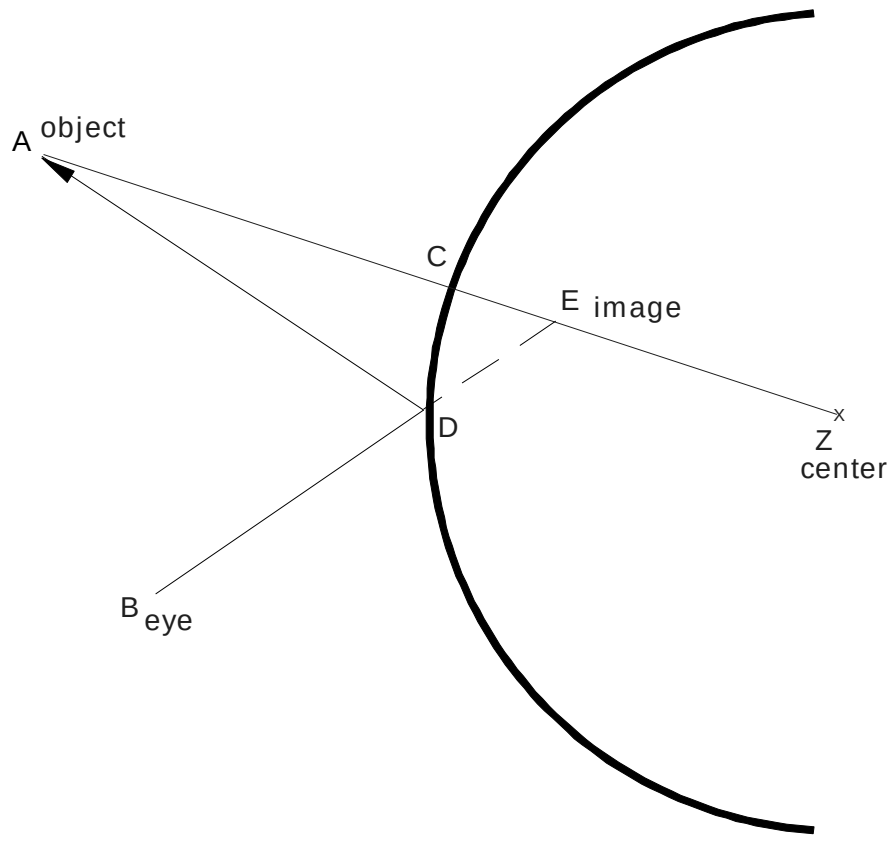

Figure 2

Euclid’s *Catoptrics*, Proposition 18 (Fig. 3)
In a concave spherical mirror the image (E) is seen at the intersection of the cathetus (perpendicular) (AD) drawn from the object (A) through the center of the sphere (Z) and the incident ray (BC).

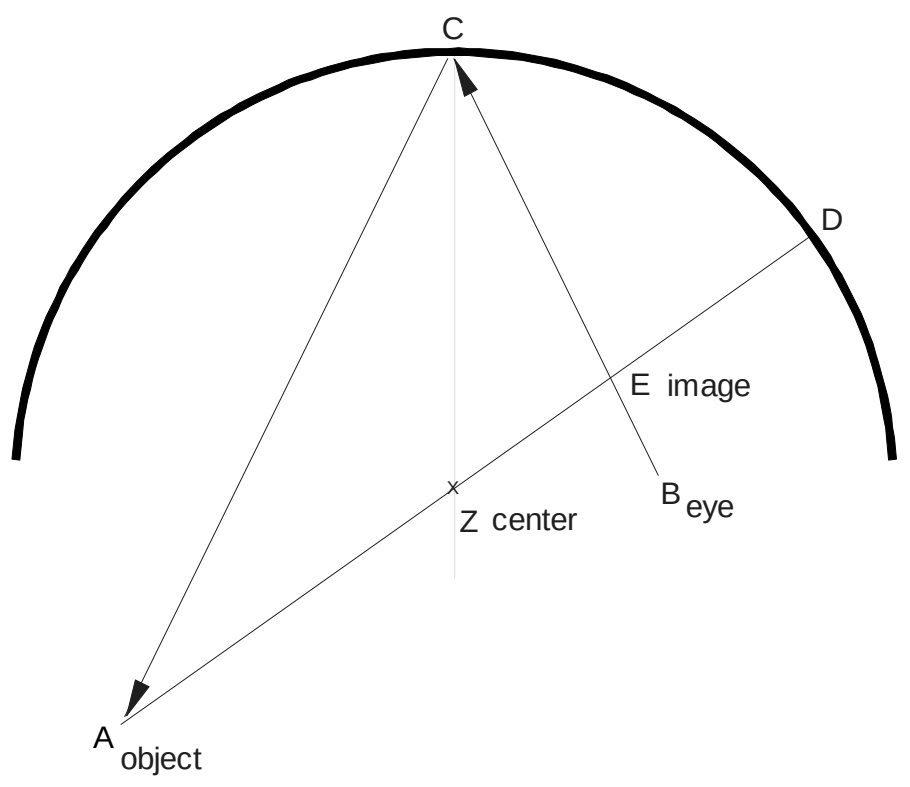

Figure 3

Independent of theories of light radiation, that is, intromission and extramission, Euclid's scheme for locating the image as presented in Propositions 16 and 17 worked well enough with plane (Fig. 4, right) and spherical convex mirrors (Fig. 4, left). Though the image changes its magnitude in convex mirror, in both cases the images appear behind the mirror in locations just as predicted by the cathetus principle.

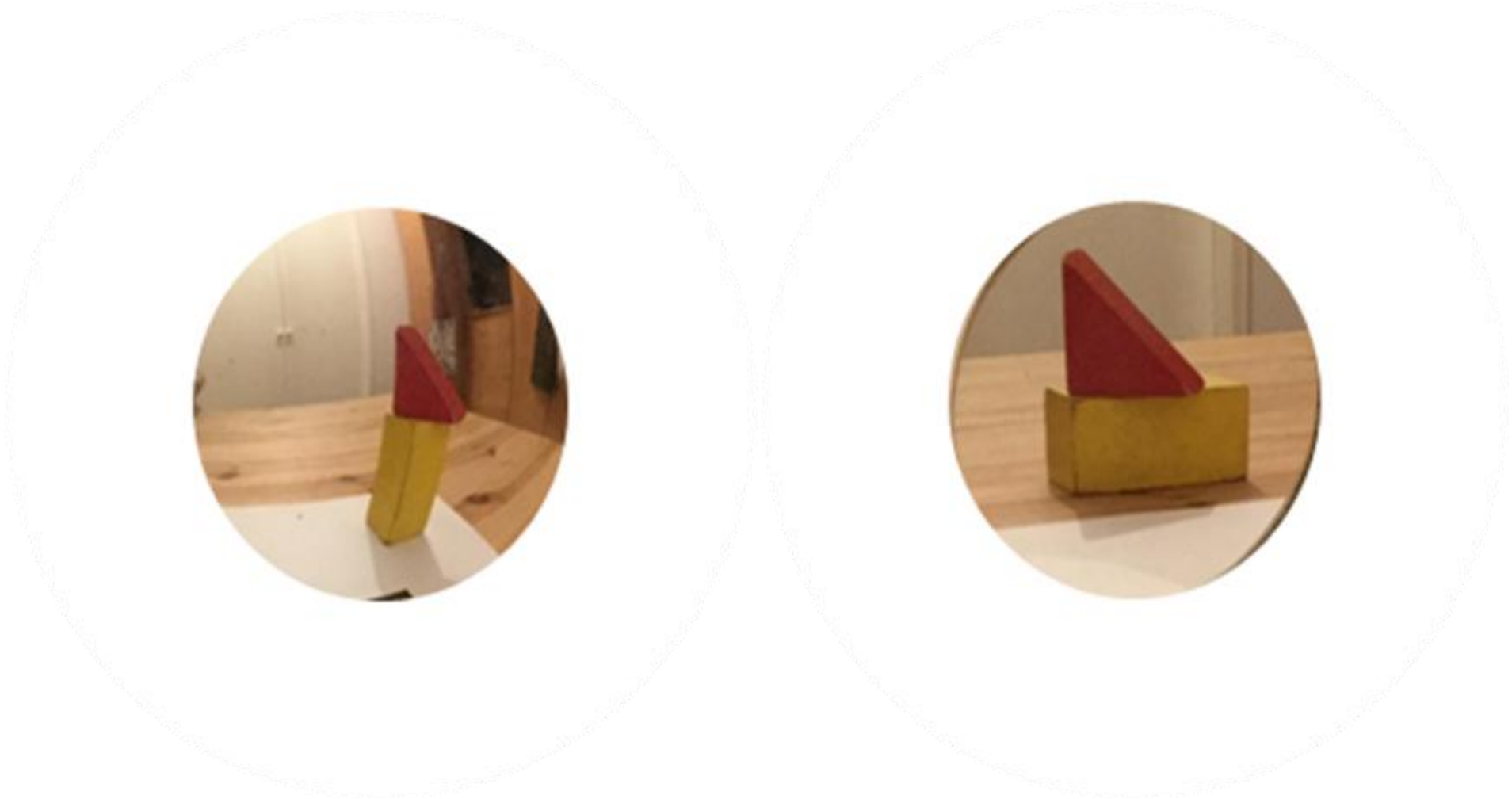

Figure 4

However, according to the very same cathetus principle, the image reflected by concave spherical mirror can be seen in several locations: behind the observer's head, in the eye, on the surface of the mirror, behind the mirror, or there may be no image at all. Moreover, at various distances of the object and the eye from the mirror, the image changes its apparent magnitude, and at a certain point (point of inversion), associated with the mirror's center of curvature, it becomes blurred and turned upside down.

Euclid's Proposition 18 thus fell short to account for these image locations. The questions, how many possible reflections there might be, and how these reflections relate to the positions of the eye, the visible object, and the image, Euclid left unanswered.

There is another important aspect of the physical properties of concave spherical mirrors that should be recalled. It was well known in Antiquity that concave spherical mirrors have the capacity to ignite fire. In Proposition 30 of *Catoptrics* (Euclid, 1537, pp. 86–87) Euclid postulated that if a concave spherical mirror is placed against the sun (Fig. 5), the rays reflected from all the points on its surface, will fall together through the center of curvature (F) where the rays become hot and ignite fire.[6]

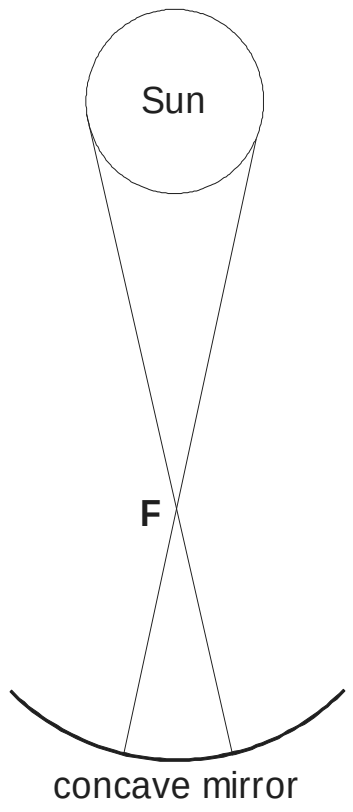

Figure 5

This is the background against which Ptolemy conducted his optical studies.

## 3. Ptolemy

### *3.1. What are the historical facts?*

1. Ptolemy wrote the *Optics* toward the end of his life, perhaps as late as A.D. 165.
2. Ptolemy's most noticeable sources were Euclid's *Optics*, *Catoptrics*, and the *Elements* (c. 300 B.C.). Possible additional sources were Archimedes' *On the Sphere and Cylinder* (C. 250 B.C.), and Hero of Alexandria's lost commentary on the *Elements* and *Catoptrics* (1st century A.D.?).
3. Ptolemy sought to account for visual perception. Like Euclid, his theory of vision is based on the visual ray emitted from the eye.
4. Reduced to a center point of sight the eye is directly linked by the visual ray to a point object. As long as this visual link remains uniform the object is seen as it actually exists in space. For Ptolemy, any deviation from this pattern of vision constitutes an anomaly.
5. The two major anomalies are reflection and refraction, they both arise when the visual ray strikes an optical interface and its path is diverted. In both cases, however, the appearance in the eye is not the object as it actually exists in the physical space, but a displaced image (Smith, 1996, pp. 19–20).
6. An anomaly is thus resolved when image location is perfectly defined with respect to both, the eye and the object.
7. Nowhere in his *Optics* Ptolemy discussed the physical properties of burning

[6] See Cardano, 1580, p. 165; Forrester, 2013, p. 244. For Euclid's discussion of reflection in a concave spherical mirror, see Knorr, 1994, pp. 16–18; Smith, 1999, pp. 84–86. See also Lindberg, 1970, Proposition 17, pp. 169–170, and Proposition 55, p. 209; Lindberg, 1983, pp. 273–277.

mirrors.

*3.2. Ptolemy's theoretical scheme*

Ptolemy's analysis of image location and distortions in plane and convex spherical mirrors was straightforward, as he followed Euclid's geometrical scheme. The generated images appear behind the mirrors, where the intersection of the cathetus and the extension of the incident visual ray must always lie.

But in his account of concave spherical mirror Ptolemy breaks from the Euclidian scheme. Depending on the position of the eye and the visible object, Ptolemy analyzed 5 cases, where the intersection of the visual ray and the cathetus define the location of the image which (Fig. 6),

1. lies behind the mirror (objects O1, O2, and images I 1, I 2);
2. there is no image at all since the incident ray and the cathetus never meet (object O3);
3. lies between the mirror and the eye (object O6, and image I 6);
4. lies at the eye (object O5, and image I 5);
5. lies behind the eye (object O4, and image I 4).

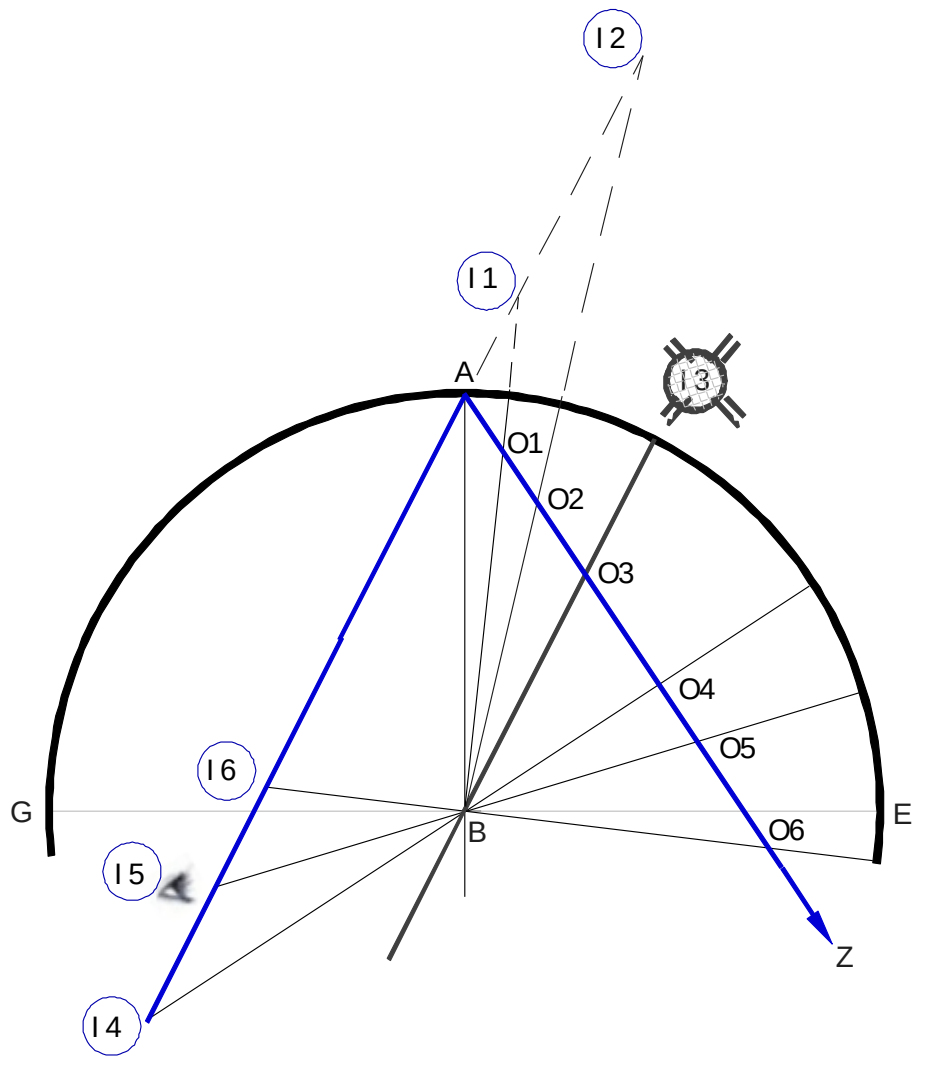

Figure 6

For the sake of brevity we will only look at Ptolemy's analysis of cases 1, 2, and 5.[7]

**Case 1** The image lies behind the mirror (Fig. 7):
Let GAE be a concave spherical mirror with center point B. With the eye placed at L, let T represent the visible object O2. Let the path of the reflected visual ray be assumed to occur at equal angles along LA and AZ. Then, from point B, let line BT be drawn longer than segment TA, and let line BZ be drawn shorter than line AZ. Finally, let the imaginable line drawn from the eye (L) to the visible object (T) fall, like LZ, between the center of the sphere B and the mirror. We say, then, that the extension of the cathetus BT intersects the extension of ray LA behind the mirror at point O, where the image I 2 is located.

[7] For Ptolemy's account of cases 1, 2 and 5, see Smith, 1996, Bk. 4, Theorem 22, pp. 192–193.

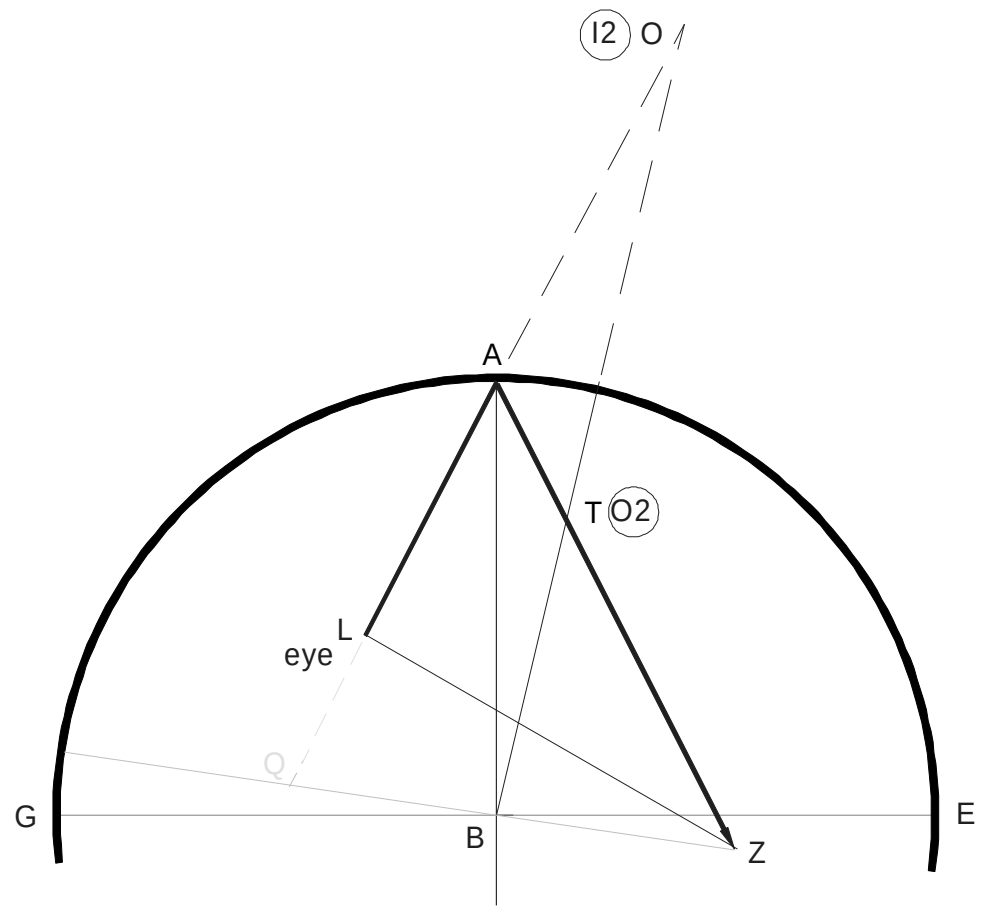

Figure 7

**Case 2** There is no image at all since the incident ray and the cathetus never meet (Fig. 8):

Let GAE be a concave spherical mirror with center point B. With the eye placed at L, let H represent the visible object O3. Let the path of the reflected visual ray be assumed to occur at equal angles along LA and AZ. Then, from point B, let line BH be drawn equal to segment HA, and let line BZ be drawn shorter than line AZ. Finally, let the imaginable line drawn from the eye (L) to the visible object (H) fall, like LZ, between the center of the sphere B and the mirror. We say, then, that BH is parallel to LA. Thus the intersection that defines the image is indeterminate, that is, the incident ray LA and the cathetus BH will never meet.

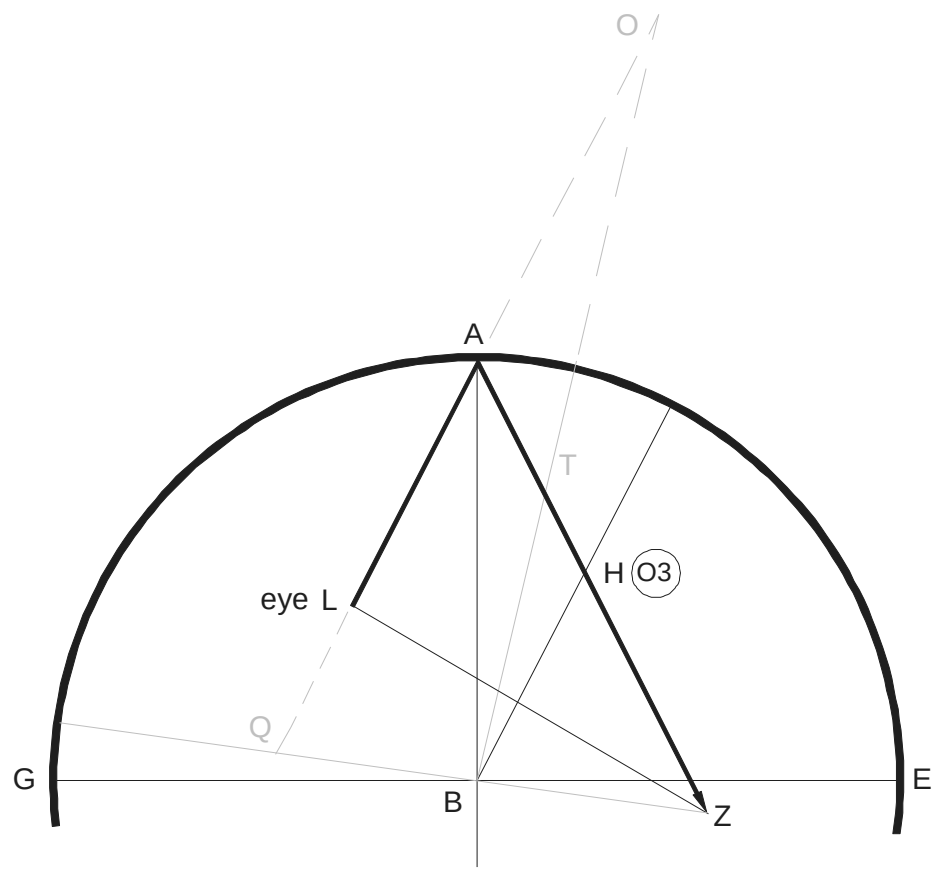

Figure 8

**Case 5** the image lies behind the eye (Fig. 9):

Let GAE be a concave spherical mirror with center point B. With the eye placed at L, and let K represent the visible object O4. Let the path of the reflected visual ray be assumed to occur at equal angles along LA and AZ. Finally, let the imaginable line drawn from the eye (L) to the visible object (K) fall, like LZ, between the center of the sphere B and the mirror. We say, then, that the cathetus KBQ intersects the extension of the ray LA at point Q behind the eye of the observer where the image I 4 is located.

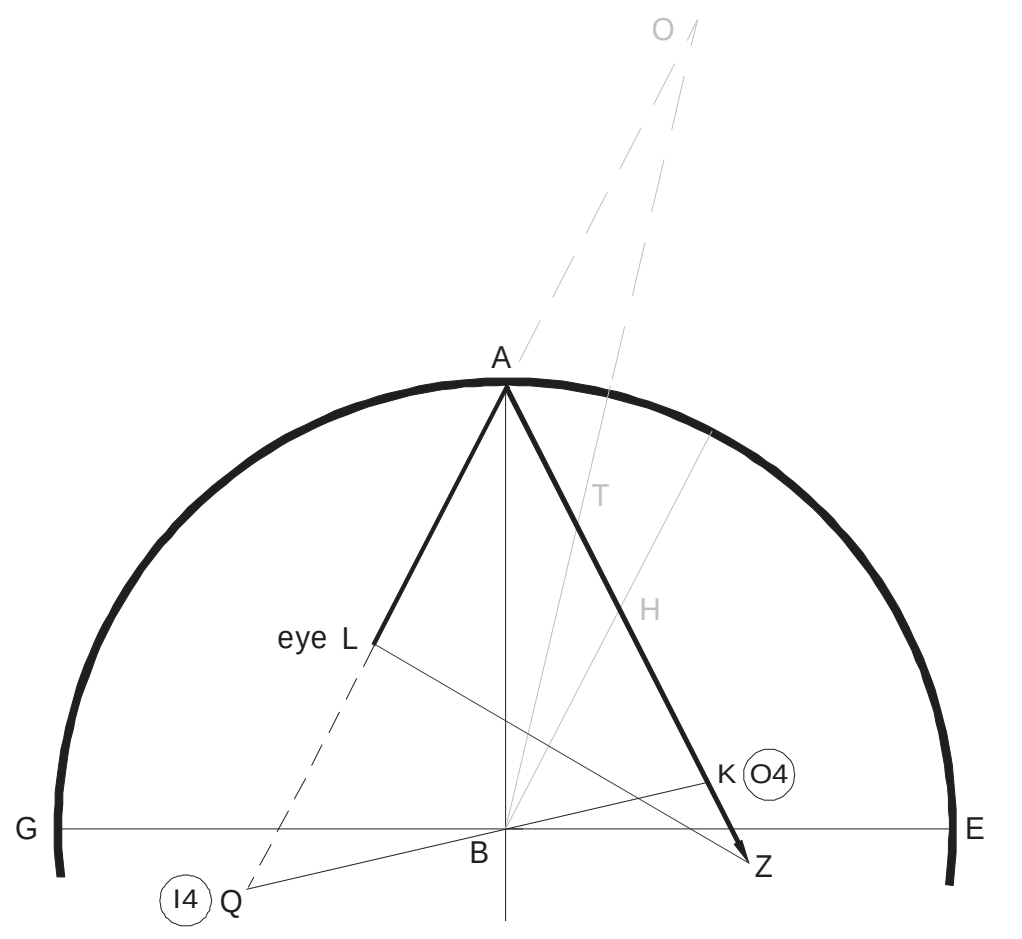

Figure 9

*3.3. Ptolemy's observational practice*

Ptolemy thought that it would be appropriate to examine the results of each one of the image locations he discussed in his theoretical analysis (Smith, 1996, Bk. 4, pp. 194–195). For that purpose Ptolemy devised an instrument for exhibiting the phenomena of reflection in mirrors. The instrument was made of a bronze disk of moderate radius (AD) size (Fig. 10), divided into four quadrants.[8] Each quadrant was subdivided into one degree increments. On a rotating arm on the disc, he set a sighting device (dioptra) that pivots around the center (A) of the disk through which, with either of the observer's eyes placed at point L, one can view objects which are placed on the disk. Three different kinds of mirrors (plane, concave, and convex with radii AB equal to AD) made of polished metal strips could be placed upright at the center of the disk. The dioptra is then rotated along the edge of the disk (GBE) until the line of sight, reflected through the concave spherical mirror TAK, is fixed so an object placed on the disk, at point M, could be seen. Now the angles of reflection (denoted by arcs LB = BM) and the location of the image for each one of the cases 1, 2, and 5, in turn, could be empirically measured and compared with the predicted location according to the cathetus principle.

[8] Cohen and Drabkin, 1966, pp. 270–271; see also Smith, 1996, Bk. 3, pp. 134–135.

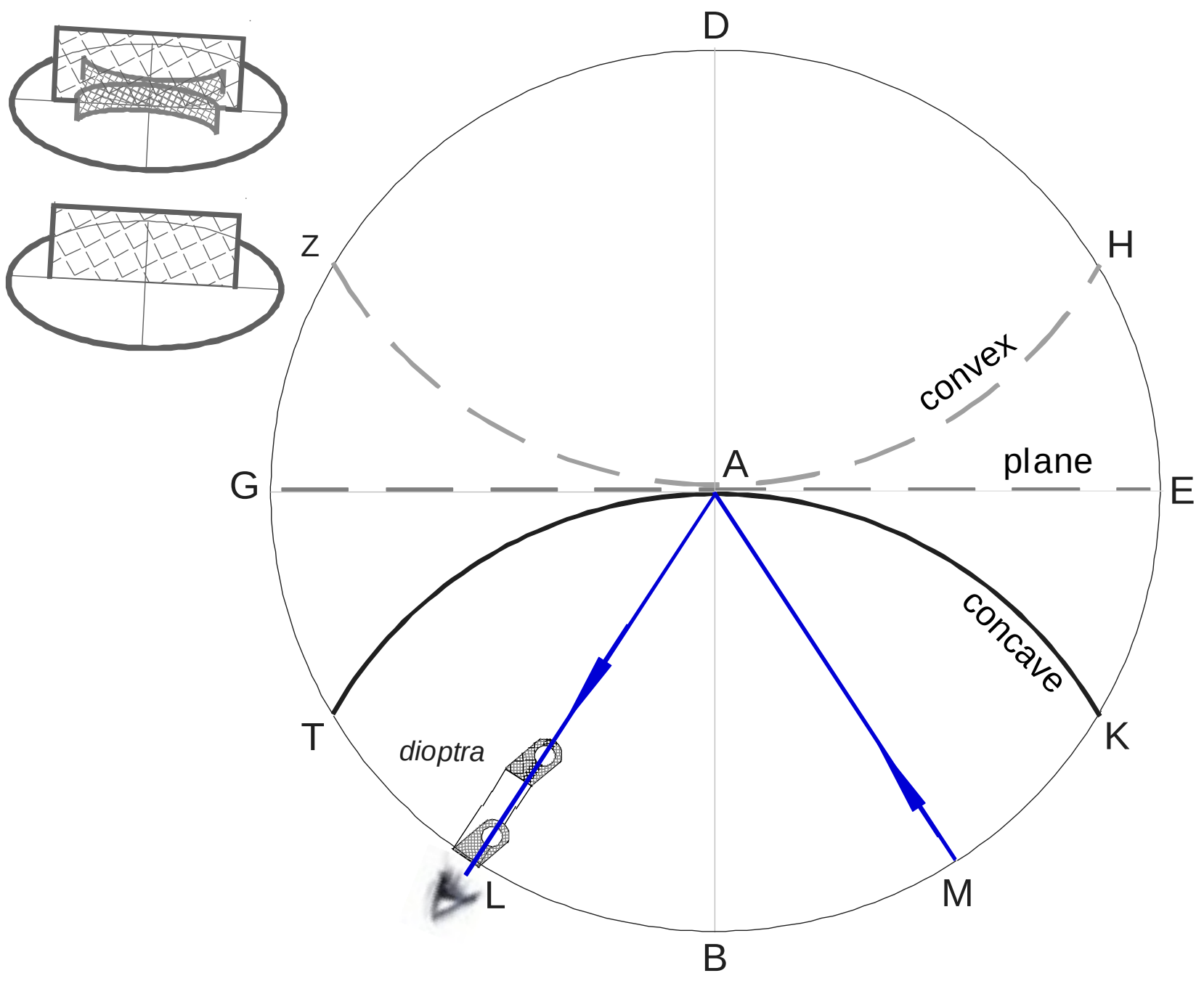

Figure 10

Ptolemy's description of the path of the visual rays reflected by a concave spherical mirror is problematic. Metal mirrors, like those which Ptolemy used, had poor optical qualities and it is questionable whether or not this arrangement was feasible.[9] Still, on the assumption that the quality of the mirrors which Ptolemy used was sufficient for exhibiting such optical phenomena, we follow Ptolemy's observational procedure: The eye is placed at point L and looks through the dioptra along LA (Fig. 11) at the concave spherical mirror TAK.[10] Given that arc LB is equal to arc BM, the image of an object placed at M, according to Ptolemy, should be seen at point A. However, Ptolemy's account is incorrect. The image of the object placed at M, as illustrated in the left upper side of Fig. 11, appears as an immensely magnified blurred patch of light rather than a clear and upright image. In effect, in this arrangement, the locations of object M and the point of inversion coincide with the result of seeing a blurred patch of light covering up the whole surface of the mirror. The images of objects O1 and O2, placed along AM, appear diminished, clear and upright behind the mirror as presented by images I 1 and I 2. All the other images are seen on the surface of the mirror at the vicinity of point A. Object O3 appears clear, upright, and a bit enlarged as illustrated by image I 3, while object O4 appears upright, more enlarged, but a bit confused as illustrated by image I 4. Object O5 appears upright and magnified but very confused as illustrated by image I 5, and object O6, which is placed further away from the edge of Ptolemy's device, is seen turned upside down, diminished and a bit confused as illustrated by image I 6.

[9] On the technological means of production and the quality of reflecting surfaces made of metal, see Schweig, 1941; Clagett, 1980, pp. 154–156; Newton, 1730 (1979), pp. 102–107; Anderson, 2007, pp. 16–32.

[10] Each simulation in this paper reproduces effects resulting from a specific setup of the mirror, the location of the object, and the placement of the eye at various distances from the mirror in reference to the point of inversion and the burning point (i.e., focal point) of the mirror. Each simulation reproduces in addition the appearance of the image in terms of its magnitude, blurriness (denoted by the hatched area), and whether it is turned reverse or upright.

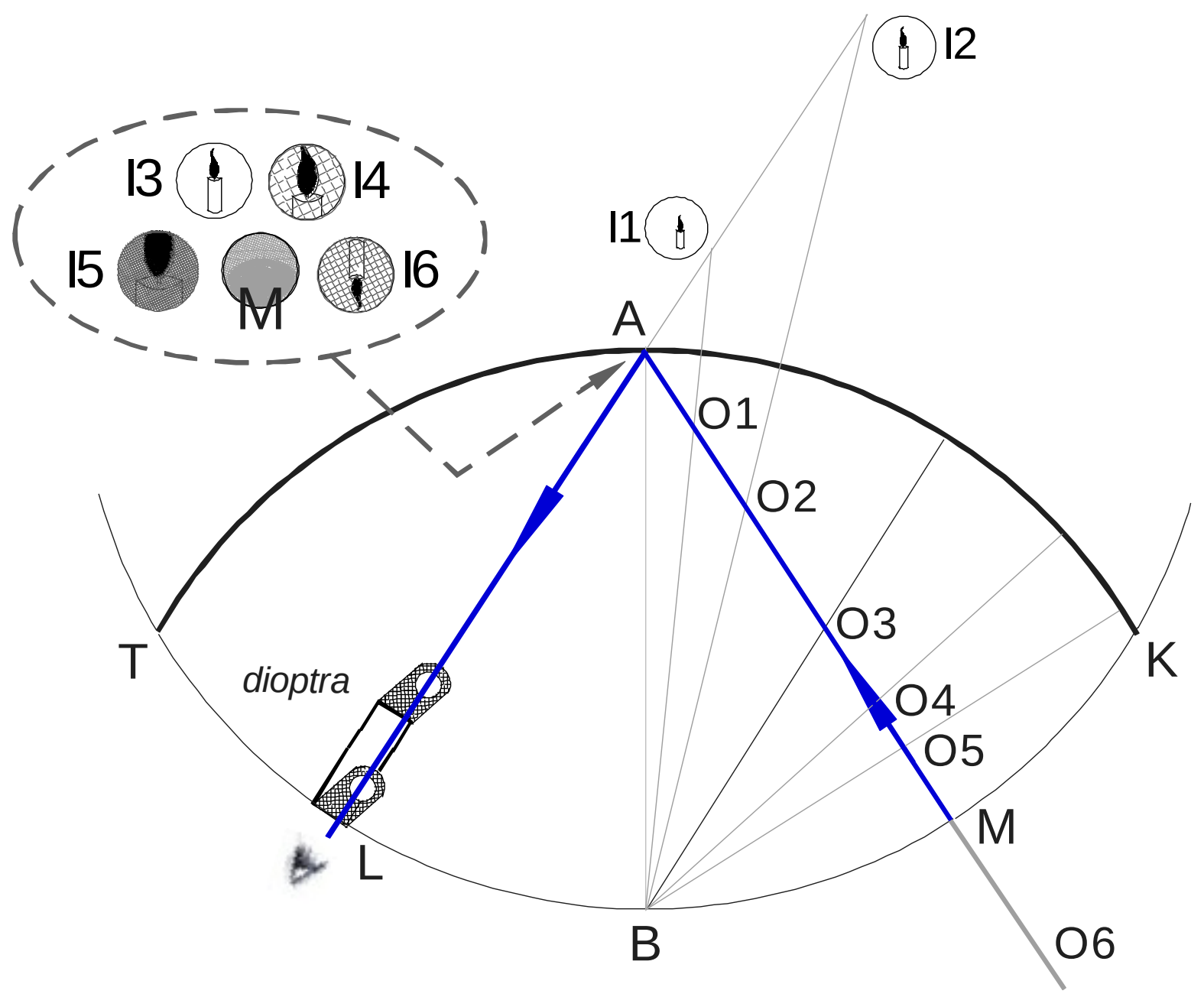

Figure 11

It is evident that the observations which Ptolemy reported for an object placed at point M cannot be reproduced. However, the observations for objects O3 and O4, placed along the visual ray MA, did enable Ptolemy to confirm the equality of the angles of incident and its reflection in concave spherical mirrors. The instrument thus exhibited the phenomenon of reflection in concave spherical mirrors; in other words, it illustrated the theoretical analysis.

While performing his theoretical analysis of cases 1, 2, and 5, Ptolemy placed the eye not at the dioptra's edge (along arc TBK in Figs. 11 and 12 respectively) but on the disk itself half the way towards the concave spherical mirror.[11] In this position (Fig. 12), the images of objects O1 and O2, placed along AM, appear diminished clear and upright behind the mirror as illustrated by images I 1 and I 2. All the other images are seen on the surface of the mirror at the vicinity of point A. Object O3 appears clear, upright, and a bit enlarged as illustrated by image I 3, while object O4 appears upright, somewhat enlarger, but a bit confused as illustrated by image I 4. Object O5 appears upright and a bit more enlarged but very confused as illustrated by image I 5. Object M, placed on the edge of the instrument appears a bit more enlarged but very confused as illustrated by the image of M, while object O6, which is placed further away from the edge of Ptolemy's device, is appeared upright, magnified and can hardly be seen as illustrated by image I 6. In this arrangement, however, the point of inversion changed its location. While in the previous arrangement the point of inversion coincides with the object placed at M, here the point of inversion appears only if the object is placed further away from the mirror behind the location of object

[11] Placing the eye at such a position exerts great inconvenience in performing the observational task.

O6. The lack of reference to this fact suggests that, like in the previous arrangement, the instrument and the procedure which Ptolemy pursued were merely for illustrative purpose; in essence, this practice was not experimental.

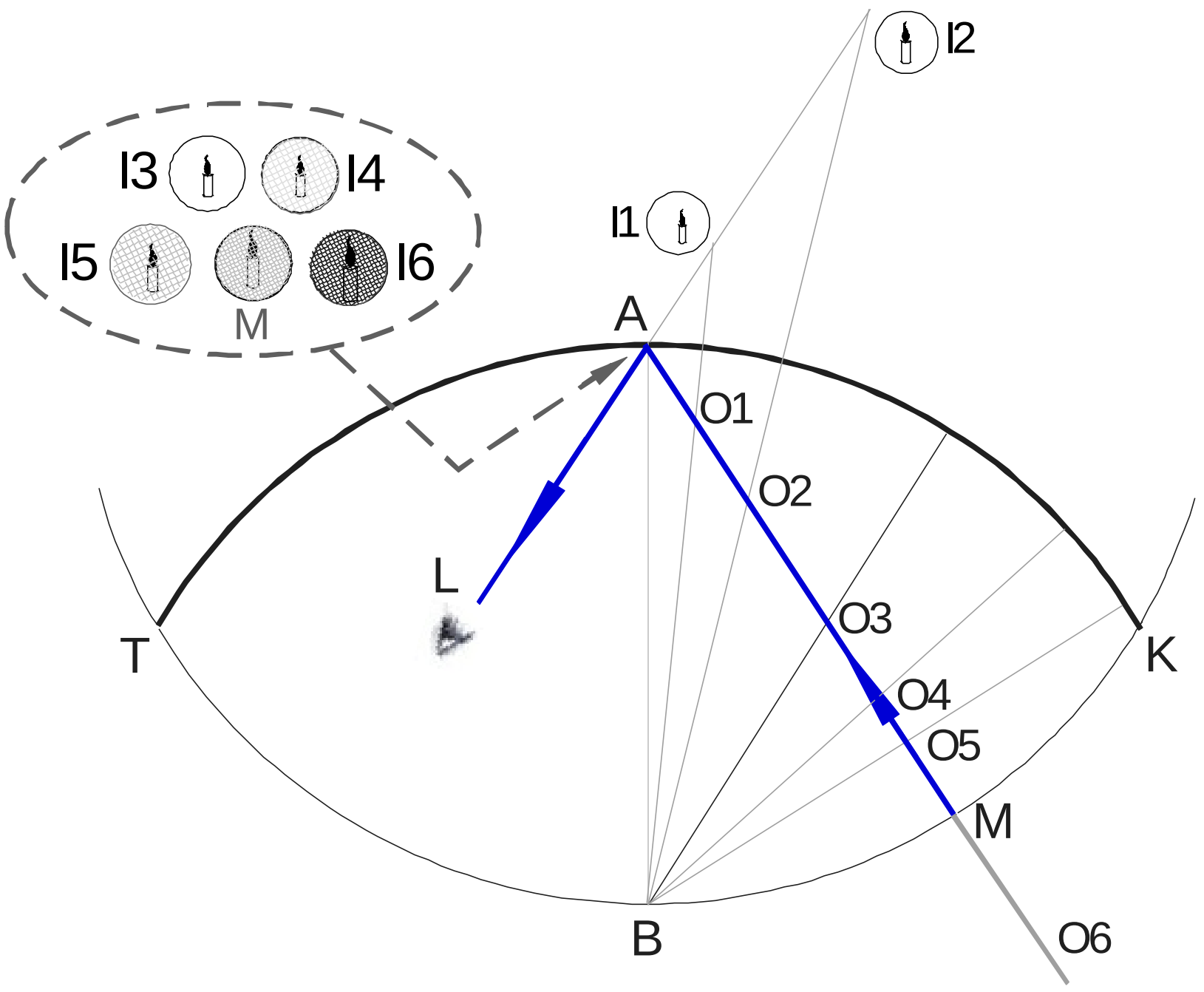

Figure 12

We apply the simulations of cases 1, 2, and 5, to illustrate the optical phenomena Ptolemy studied in his observations with concave spherical mirrors.

**Case 1** (Fig. 13)
When the object O2 is placed at T and the eye at L, the image appears behind the mirror at point O, where the extension of the incident ray LA intersects the extension of the cathetus BT. Thus, image I 2, related to this object is seen as predicted behind the mirror.

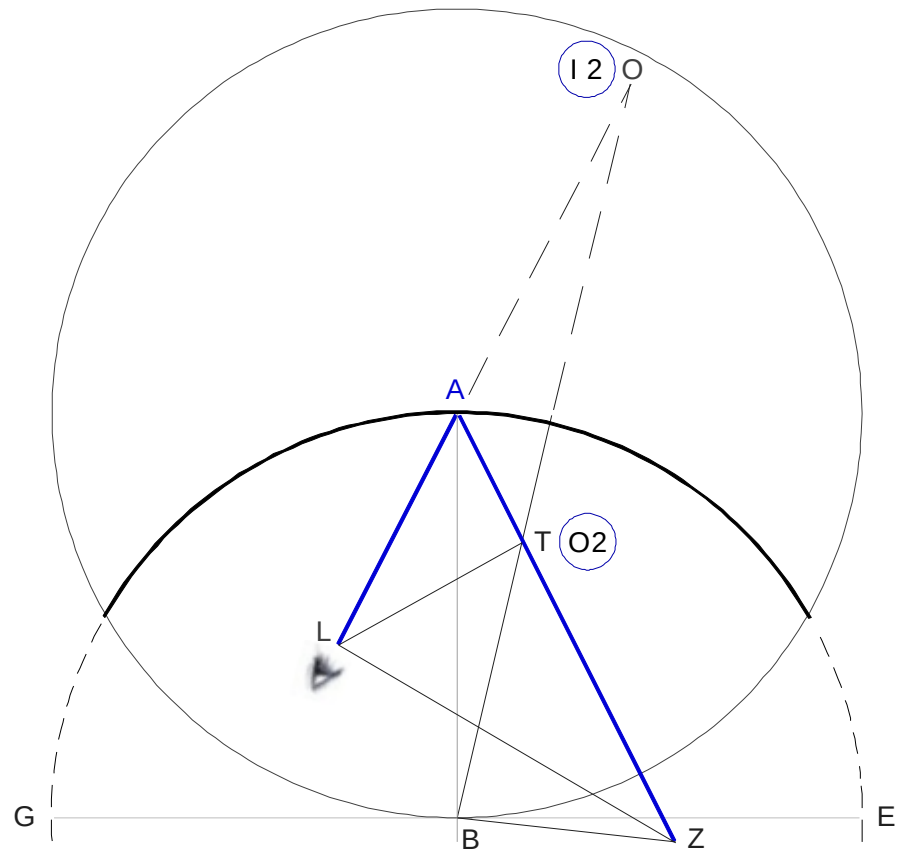

Figure 13

**Case 2** (Fig. 14)
When the object O3 is placed at H and the eye at L, the incident ray LA and the cathetus BH are parallel. The intersection is undetermined and thus nothing can be seen. In contrast, Ptolemy's observations show that the image I 3 appears to lie in front of the mirror at the vicinity of point A.

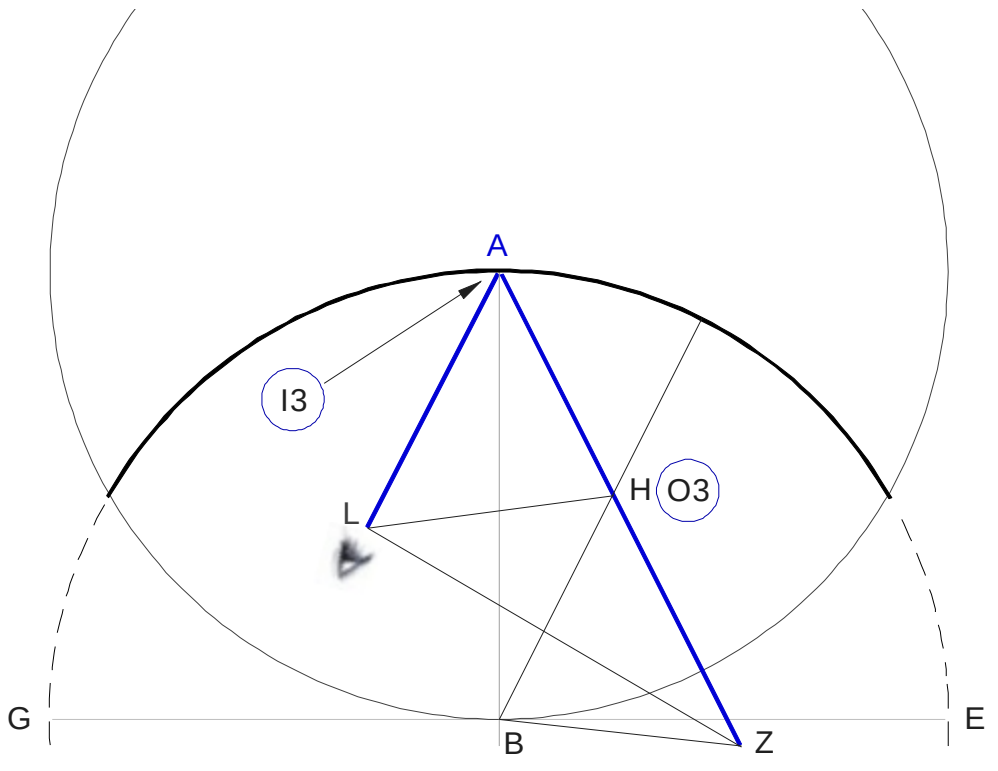

Figure 14

**Case 5** (Fig. 15)
When the object O4 is placed at K and the eye at L, the extension of incident ray AL and the cathetus KB intersects behind the eye. The image should be located behind the viewpoint, but this is impossible. In contrast, Ptolemy's observations show that the image I 4 appears in front of the mirror.

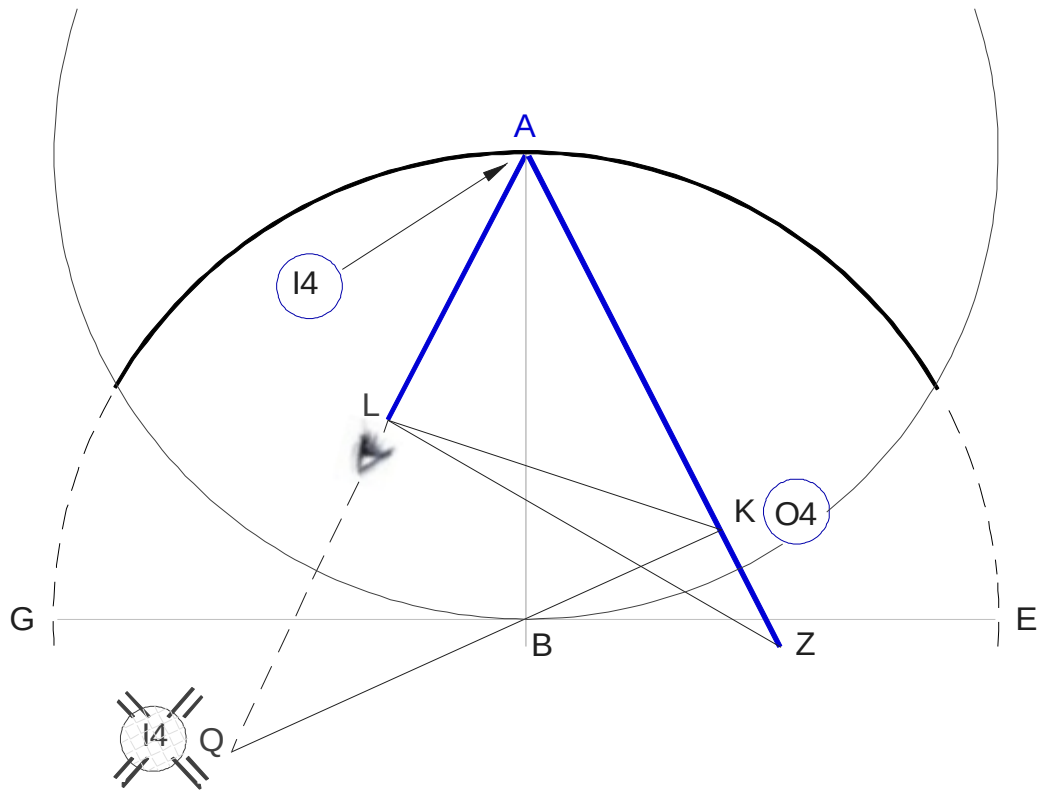

Figure 15

In sum, as shown in Fig. 16, only the images of objects O1 and O2 appear behind the mirror as predicted by the cathetus principle. In contrast, the location of images I 3, I 4, I 5, and I 6 which are related to objects O3, O4, O5, and O6 respectively, do not conform to Ptolemy's theoretical analysis for they all appear in front of the mirror at the vicinity of point A. There is then a fault in this analysis of which Ptolemy was aware. He argued that although the image may be out of sight, it is not out of mind; thus, any disruption in the appearance of an image generated by a concave spherical mirror will be perceptually adjusted by the visual faculty so that the final image is seen as similar in kind to its generating object.[12] In effect Ptolemy introduced the mind as an active agent, an integral element of the optical scheme, for rectifying the discrepancy between the theoretical and the actual locations of the image as seen at the eye. The visual faculty was thus part and parcel of the physics of optical elements. The perceiving agent was an essential part of the phenomenon; vision was not external to the phenomenon of reflection.

[12] Smith, 1996, Bk. 4, p. 194, also see p. 197.

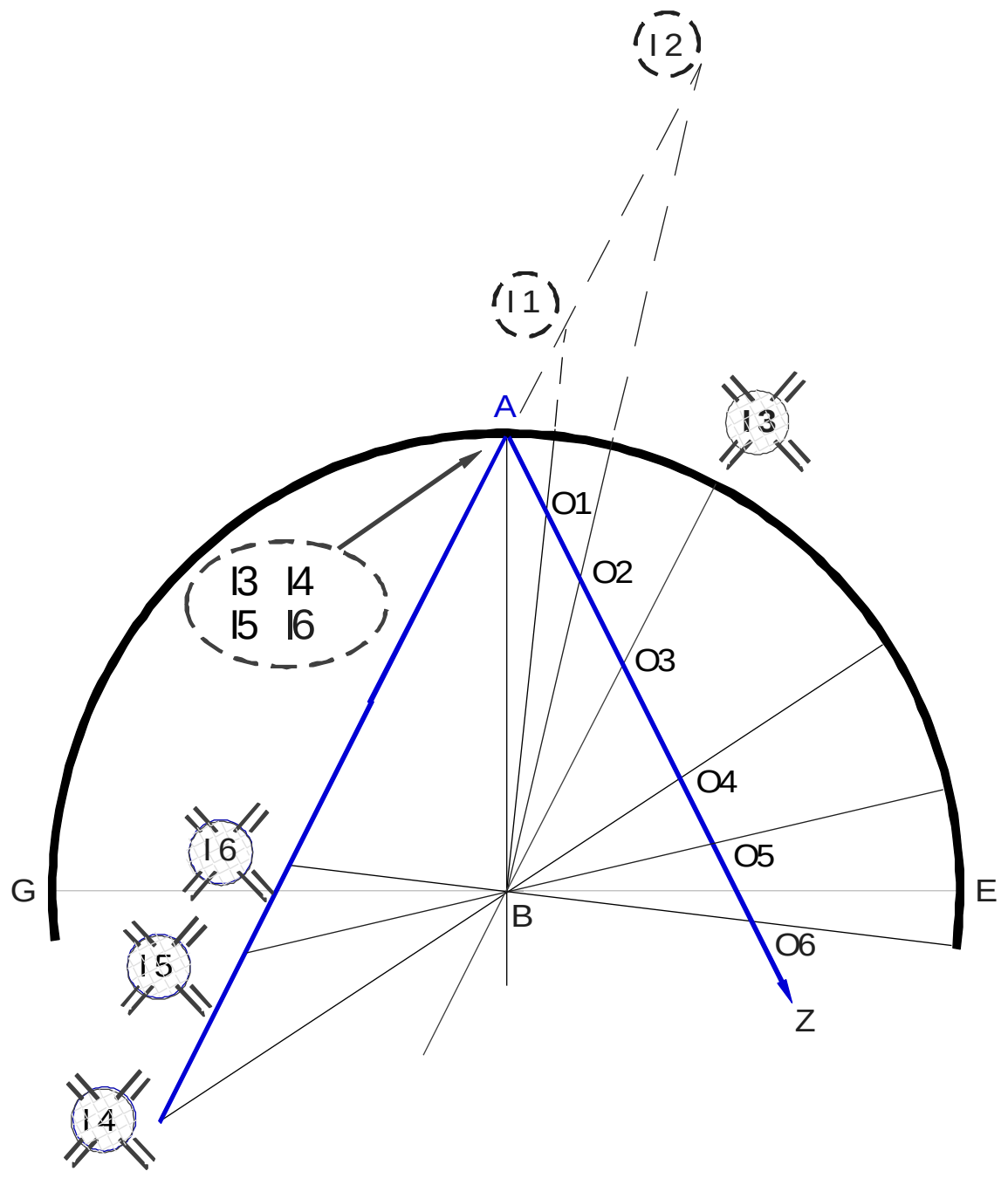

Figure 16

## 4. Della Porta

### *4.1. What are the historical facts?*

In Bk. 17, Ch. 4 of *Magia naturalis Libri XX* (1589) Della Porta reported his experiments on the functioning of concave spherical mirrors arranged in various setups.[13] As noted earlier, since the time of Euclid it was commonly acknowledged that when a concave spherical mirror is set against the sun, the reflected rays are directed to the center of the mirror onto a single point (*vnum punctum*) where fire is kindled.[14] Della Porta experimentally explored wide variety of visual effects that could be generated by concave spherical mirrors. He confirmed the distinction between two optical points:
(1) *Punctum inversionis*: the point of inversion in reference to the position where the image, seen reflected in concave spherical mirrors (and convex lenses, subject to the phenomenon of refraction), achieves its greatest magnification just before it turns upside down and appears blurred, as the eye is moved further away from the optical element; and,
(2) *Punctum incensionis*: the point of burning in reference to the physical point at which light rays are concentrated by reflection in concave spherical mirrors (and in refractive elements such as convex lenses).
Two issues are worthy of note here. In the first place, Della Porta considered reflection and refraction reciprocal phenomena, that is, he treated the two distinct phenomena as complementary (Della Porta, 1593, Bk. 2, p. 35). This is a fundamental innovation. Secondly, in contrast to opticians and practitioners of the time who distinguished between the point of inversion and the point of burning but did not spatially separate them, Della Porta took the two points to occupy different spatial locations.[15]

[13] Della Porta, 1589, Bk. 17: 264–265.
[14] See fn. 6 above. See also Cardano, 1580, Bk. 4: 164–166; Forrester, 2013, Bk. 4: 244, 246.

We analyze Della Porta's study of reflection in concave spherical mirrors assuming the following historiographical presupposition: Bk. 17 of *Magia naturalis* and *De refractione* should be read as one body of optical research. Della Porta made ample cross-references in the two books. We first review the historical facts, as they appear in Bk. 17, Ch. 4 of *Magia naturalis*:[16]

1. From the outset of Bk. 17 and up to the end of Ch. 3, Della Porta conducted experiments with single as well as arrays of plane mirrors.
2. In Ch. 4 Della Porta turned to study the optical effects of concave spherical mirrors. Right from the beginning of the chapter he emphasized the need to know the position of the point of inversion, which he then associated with the point where the reflected rays of the sun are united.
3. Della Porta experimented on how the functioning of concave spherical mirrors, in various setups of the eye and the object, affects the appearance of the image at the eye.
4. At the end of Ch. 4 Della Porta introduced the point of burning into his analysis, namely, the point where the converging rays create the most powerful brightness and heating capacity to ignite fire. From this juncture to the end of Bk. 17 Della Porta did not discuss any more the point of inversion; he referred only to the point of burning.
5. At the end of Ch. 6 Della Porta made the first of 7 references in which the reader is encouraged to consult his other optical book, that is, *De refractione.*[17]
6. In Ch. 22 of Bk. 17 Della Porta provided technical instructions of how plane and spherical mirrors of good quality are produced.[18] Della Porta's account was not merely theoretical; he owned a workshop for producing optical elements.[19] This is evident in Francesco Fontana's (1580–c. 1656) attempts to get possession of the instruments for lens

[15] Smith, 2015, p. 344; Dupré, 2005, p. 152.
[16] For background discussion on Della Porta's scientific enterprise, see the introduction and conclusion in Borrelli et al, 2017, pp. 5—10, 201—205; Zik & Hon, 2017. On the scientific culture of Renaissance Naples see, Eamon, 2017, pp. 11—38.
[17] Della Porta, 1589, Bk. 17, pp 267: "In nostris in opticis fusius declaratum est."
[18] Della Porta, 1589, Bk. 17, Ch. 22, pp. 279. Note that Della Porta's technical information given in chapters 17–21, and 23 of Bk. 17, encompassed detailed instructions for the production of parabolic mirrors, optical elements of cylindrical shape, spectacles lenses, and mirrors made of back coated glass or metal, see Della Porta, 1589, Bk. 17, pp. 275–280.
[19] On Della Porta and his optical experience in Venice, see Reeves, 2008, pp. 70–78.

making left after the death of Della Porta in 1615.[20]

The facts above bring us to make two fundamental claims: (1) Della Porta experimented on concave spherical mirrors which facilitated quantitative demonstration of the phenomenon of reflection, and (2) Della Porta developed a novel theory of reflection in concave spherical mirrors which eliminated perceptual considerations from optics and considered only geometrical-physical aspects. Clarification of these two claims demands of us to change the order of our analysis. While in Ptolemy's case we began with theory and then moved on to observations, here in Della Porta's case we begin with observations, indeed experiments, and only then move on to theory. We claim that Della Porta experimented on reflection in concave spherical mirrors; he did not consider his observations mere illustrations of the phenomenon; rather, he searched for regularities of the phenomenon of reflection which could be formulated quantitatively.

### *4.2. Della Porta's experimental practice*

In Bk. 17, Chap. 4 of *Magia naturalis* titled, *On the different effects of concave mirror*, Della Porta experimented on visual effects which are produced by concave spherical mirror (Della Porta, 1589, Bk. 17, pp. 264─265).[21] Della Porta adhered to the old notion of reflection in concave spherical mirror, yet he emphasized the importance of the location of the point of inversion, that is, the place where the image is seen turned upside down. In order to find the place of that point one has to set the mirror against the sun and where the reflected rays are united there the point of inversion is located.[22] But to see things magnified, one should set one's head below the point of inversion, closer to the surface of the mirror, where one could see one's face huge and monstrous and one's finger as great as one's arm. At that point women could pull hairs of their eyebrow for they are seen as large as finger.[23] Della Porta's argument can be demonstrated in the following simulation (Fig. 17): The appearance of the image when the eye and the object (e.g., a candle) are being gradually moved further away from the mirror. Near the mirror the image is seen upright and diminished. Further away from the mirror the image is gradually getting magnified but still seen clear. The image is getting blurred as it came nearer to the point of inversion. At the center of curvature, where the point of inversion is located, the image is most magnified, reversed, and blurred. Then, with the increasing range, the image is seen reversed, less blurred, and diminished.

---

[20] Crasso, 1666, Vol. 2, p. 297.

[21] For a comprehensive analysis of Della Porta's experiments with concave spherical mirrors, see Zik & Hon, 2017, pp. 43─50. For the sake of brevity, we address in this paper only part of Della Porta's experimental demonstrations.

[22] In effect this is the place where the point of burning (focal point) of the mirror is located.

[23] Della Porta, 1589, Bk. 17, p. 264.

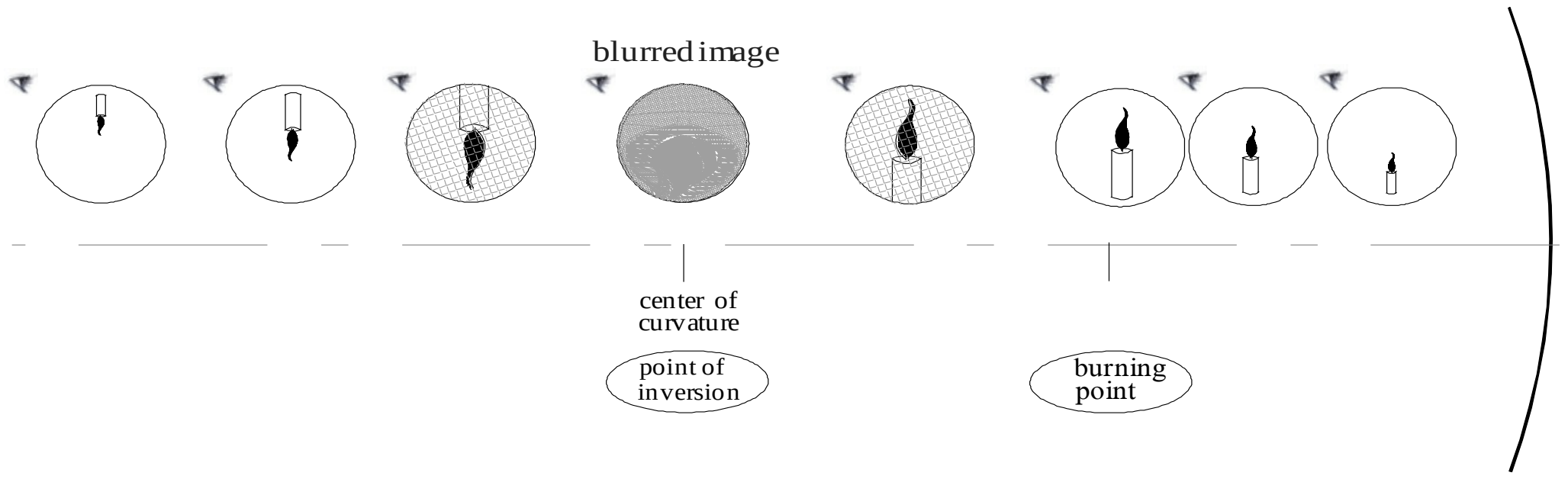

Figure 17

While exploring the capacities of concave spherical mirror's to project heat, cold, voice, or drawing the projection of an image on a board, Della Porta could not miss the visual appearances associated with the center of curvature of the mirror (Fig. 18).[24] When the object and a screen are placed at the center of curvature, the image is depicted on the screen inverted at the same magnitude. If the eye is replacing the screen the image is seen most magnified, inverted, and blurred.

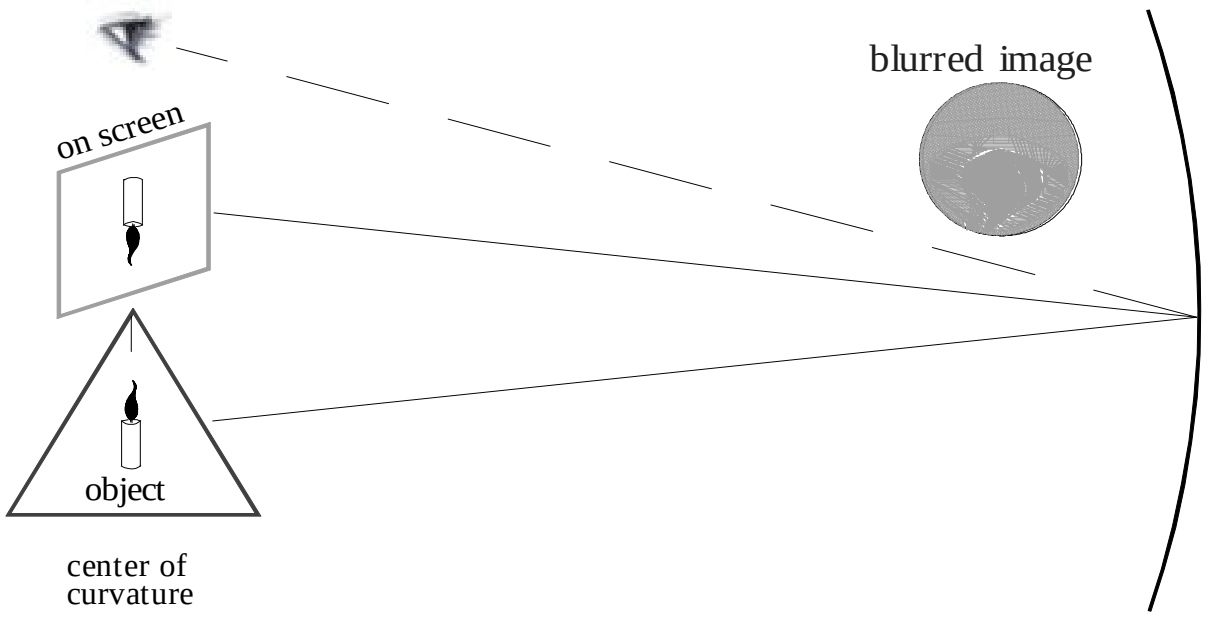

Figure 18

Della Porta then explored the appearances of the image when the eye is moved further away from the center of curvature (Fig. 19)**:** The image is seen inverted and less magnified and the picture of a thing (object) shall be farther stretched forth. If one moves the eye at that place to the right or left-hand toward the edges of the mirror, the picture of things would be seen stretched forth about the mirror surface at the point where the cathetus touches the line of reflection. In so doing many strange wonders, which few persons were able to reproduce, may be observed.[25] But when the eye is fixed further away from the center of curvature, and the object is being gradually moved away from the mirror, the point of inversion moves and it is not located any more at the center of curvature; it changes its location and movies closer to the mirror. The image is seen upright and diminished near the mirror and getting larger and blurred towards the point of inversion. Further away the image is reversed, getting clearer, and diminished.

[24] Della Porta, 1589, Bk. 17, pp. 264, 266.
[25] Della Porta, 1589, Bk. 17, p. 264.

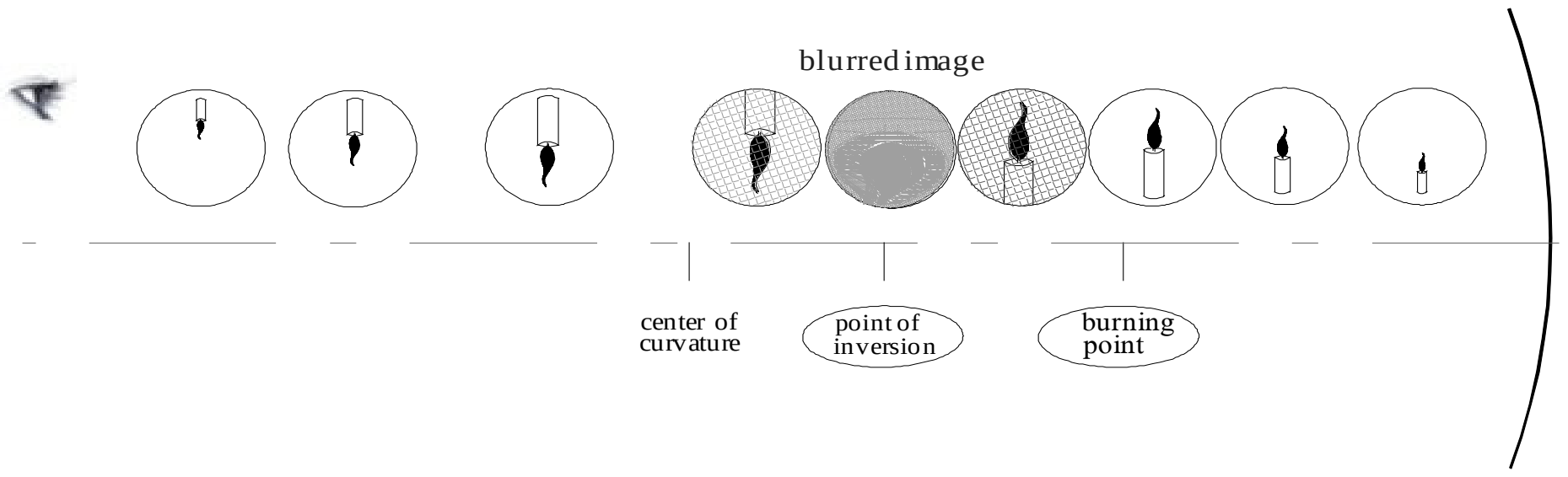

Figure 19

The point of inversion changes its location again when the object is placed closer to the mirror and the eye is being gradually moved further away from the mirror (Fig. 20): The point of inversion is located further away behind the center of curvature. Near the mirror the image is seen diminished and upright and getting larger and more blurred towards the point of inversion. Further away the image is seen reversed and then it becomes clearer and diminished.

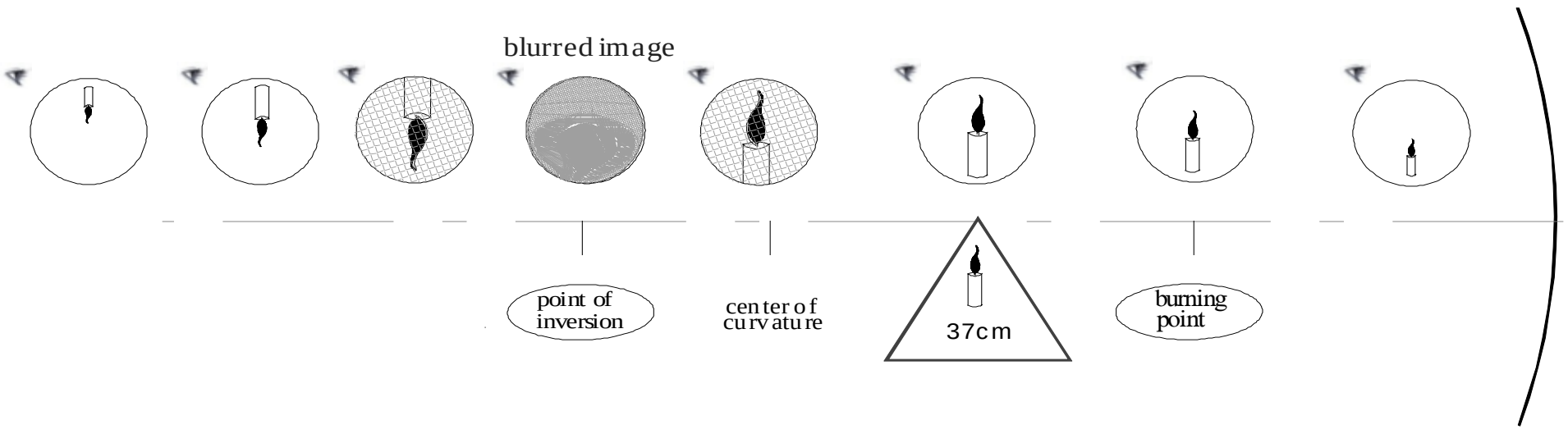

Figure 20

As reported by Della Porta, these optical experiments with concave spherical mirrors (for which we have provided several simulations), were not merely illustration of phenomena; in effect, they attest to a systematic experimental practice which yielded the following novel insight: The place of the point of inversion is not fixed. It varies according to the relative positions of the object and the eye. Della Porta emphasized that none before him had been able to demonstrate this fact.[26]

When Della Porta experimented on how letters can be read in the dark night he introduced a new reference point: To read letters in the dark night one should set a mirror against Venus or Mercury, or against a fire or light that is far off. The reflected rays will meet at the burning point (*incensionis puncto coibit*) and cast a most bright light whereby one could easily read the smallest letters.[27]

Della Porta discovered experimentally that in concave spherical mirrors the point of inversion and the burning point are not located at the same place. This discovery contradicted the received view concerning the properties of concave spherical mirrors from Antiquity up to the time of Della Porta. He associated the point of inversion with visual perception while the point of burning with physical, optical location associated with a geometrical point at which the converging rays ignite fire. Della Porta then excluded the point of inversion from his optical discussion in Bk. 17 and referred only to the point of burning. His formulation is quantitative.[28]

[26] Della Porta, 1589, Bk. 17, p. 267: "Nec leues poterunt imaginari technae, distantiam speculi magnitudine emendabis. Sat habes, qui id docere conati sunt, non nisi nugas protulere, necaliquibus ad huc compertum putarim."

[27] Della Porta, 1589, Bk. 17, p. 265.

### *4.3. Della Porta's theoretical scheme*[29]

Della Porta characterized optics as mathematical knowledge (*mathematicas scientias*) in which catoptrical experiments (*experimentae*) combined with mathematical demonstrations could provide definitive conclusions. He thought that geometrical speculations supported by observations should be considered true. He then formulated the following rules which were accepted by authors of optics (*perspectiuae authoribus accepta*): (1) Some of the rays emanating from the sun are parallel to one another; (2) The rays correspond to straight lines in geometrical demonstrations; and (3) Solar rays incident on the surface of convex, concave, or plane mirrors always form equal angles of incident and reflection.[30]

Della Porta initiated in *De refractione* geometrical study of the relation between the incident and reflected rays in concave spherical mirrors. As noted earlier, Della Porta referred the reader to the experiments with concave spherical mirrors he made in *Magia naturalis*.[31] He built his analysis on Euclid's *Elements*, Bk. IV, in which the geometrical properties of polygons inscribed in a circle are discussed. [32] Della Porta adopted Ptolemy's method of geometrical calculations and table of chords in a circle as a mathematical operation correlating the ratios between arcs and line segments. On the basis of these relations Della Porta concluded the discussion of reflection in concave spherical mirrors with a general proposition (Fig. 21a): The intersection of a perpendicular FE, dropped from the mid-line connecting the point of incident B and the center of a concave spherical mirror D, and the diameter of the mirror, always marks the point of reflection E (i.e., point of burning).[33]

---

[28] Della Porta, (1558) 1562, Bk. 4, Ch. 14: 125: "Si verò maioris sphæræ fuerit segmentum, per maiorem accendit distantiam." Note that already in the first edition of *Magiae natvralis* Della Porta had maintained that important correlation exists between the segment of the circle, that is, the region bounded by a chord and the arc subtended by the chord, of the concave spherical mirror and the distance at which it kindles fire.

[29] For a comprehensive analysis of Della Porta's theory of concave spherical mirrors, see Zik & Hon, 2017, 44, pp. 50–53.

[30] Della Porta, 1593, Bk. 2, pp. 35–36; Smith, 2019, p. 35. See also Della Porta, 1593, pp. 3–6; Smith, 2019, pp. 9–12, and Della Porta, 1589, Bk. 17, pp. 259–260. On Renaissance art, naturalism, and optics, see Smith, 2015, pp. 298–321.

[31] Della Porta, 1593, Bk. 2, p. 40; Smith, 2019, pp. 88–89: "Vt libro naturalis Magiae demonstrauimus."

[32] See Densmore, 2002, Bk. 4, pp. 83–98.

[33] Della Porta, 1593, Bk. 2, p. 41; Smith, 2019, pp. 88–91.

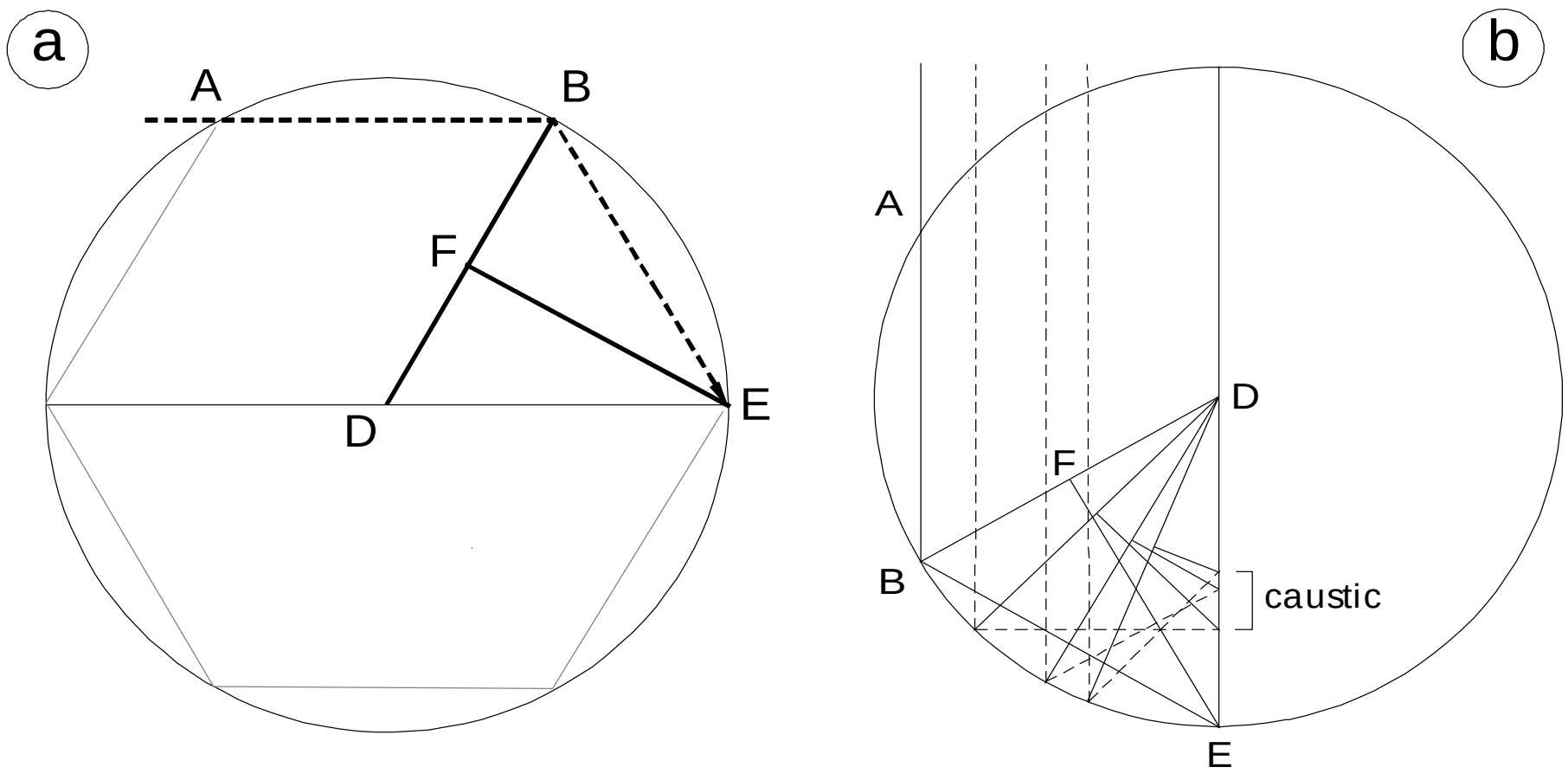

Figure 21

By applying an equilateral polygons inscribed by a circle, Della Porta established a method by which the relations among, (1) the height of the incident ray denoted by its chord, (2) the angle of reflection, and (3) the radius of the mirror, could be calculated. Accordingly, the focal length of any given concave spherical mirror could be determined in terms of the radius of the spherical mirror. One could thus establish quantitatively the caustic (i.e., spherical aberration) formed by reflection (Fig. 21b).

Hence (Fig. 22), the point of reflection for each ray, respectively, that is, point E, is located along the diameter CK from the terminus at C up to a point (E) placed at one quarter of the mirror's diameter, where the burning point of the mirror is located.

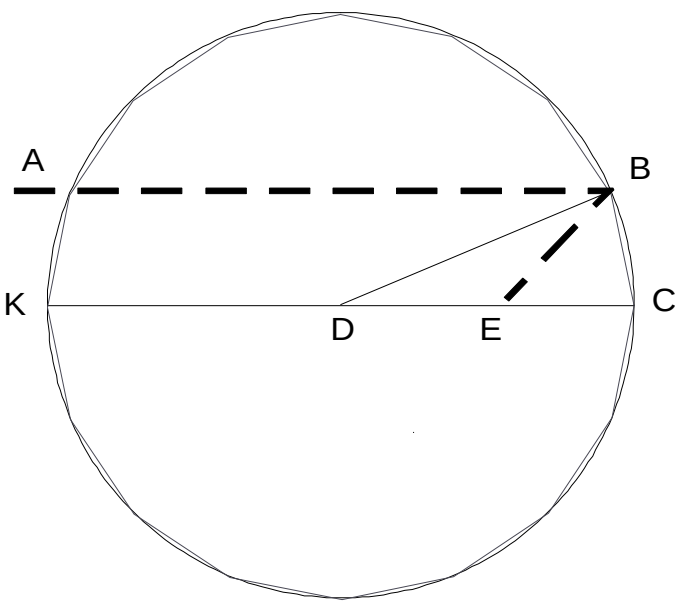

hexdecagon (16 sided polygon)

Figure 22

Della Porta determined the location of the point of burning as a function of the mirror's radius curvature; the point of burning is located on the optical axis at a distance equals to $^{1}/_{4}$ of the mirror's diameter.[34] He noted that by placing the burning point, that is, the focal point of the mirror, at one quarter of its diameter he corrected Euclid's long standing error in *Catoptrica*, where the burning point of a concave spherical mirror was placed at the center of the mirror's diameter.[35]

For this optical analysis Della Porta referred the reader to his other optical book,

[34] Della Porta, 1589, Bk. 17, p. 265, 271.
[35] Della Porta, 1593, Bk. 2, p. 39; Smith, 2019, pp. 86–87.

*De refractione optices parte libri nouem*, which he published formally in 1593.[36] In *De refractione* Della Porta explained, *inter alia*, the novel geometrical method he developed for determining the location of the burning point of concave spherical mirrors as a function of the mirror's radius curvature and the height of the incident ray.[37] This knowledge is crucial for properly setting the correct distances between the optical elements used in his optical experiments. Throughout Bk. 17 Della Porta claimed that all the experiments, either with plane or spherical mirrors and lenses, would fail if knowledge of the exact radius of the circles from which the optical elements are configured were not available. The replacement of a subjectively dependent locus, the point of inversion, with an objectively dependent locus, the point of burning, constitutes a fundamental contribution towards the technological management of sets of optical elements.

Della Porta formulated a theoretical scheme based on inferences from experiments. Using a straightedge, compass, and table of chords, Della Porta could determine geometrically optical properties of a concave spherical mirror in terms of the radius of curvature of the mirror and the height of the incident ray. Given these parameters, he could calculate the place of the focal point on the optical axis of any concave spherical mirror. It was the exclusion of visual perception from the physics of optical elements that made it possible to render optical phenomena quantitatively.

## 5. Conclusion

We are now in a position to answer the question: What is the principal difference between Ptolemy and Della Porta concerning the phenomenon of reflection in concave spherical mirror, and what can we learn from this difference about the conception of regularity in nature, specifically in optics?

The thrust of Ptolemy's study is aimed at determining the location of the image seen at the eye; Ptolemy sought to account for what the eye sees rather than for how does concave spherical mirror function. Ptolemy did not address the physics of light radiation as it encounters optical elements. His approach to reflection was that of the mathematician or, rather, the geometrician whose goal was to explain visual phenomena rigorously and comprehensively.[38] Following Euclid, Ptolemy presupposed that, like for plane and convex mirrors, the equality of the angle of incident and its reflection holds also for concave spherical mirrors. Ptolemy designed an instrumental arrangement (a dioptra) to illustrate his geometrical analysis of the phenomenon of reflection in mirrors of different shapes, namely, plane, convex, and concave. His observational procedure presupposed then the self-evident principle of reflection in concave spherical mirrors. Ptolemy was aware of the limitation of his analysis; his instrumental arrangement indicated that strict correlation between the cathetus principle and the location of the image in some specific cases related to reflection in concave spherical mirrors does not hold. Ptolemy discovered that, given the cathetus principle, the impression arising in the eye, using concave spherical mirror, sometimes do not posit the image in its expected place. The principle proved to be limited. To be sure, for Ptolemy the lack of correspondence between the theoretical analysis and the observational results with concave spherical mirrors was logically impossible. Fundamental to his visual theory was the supposition that whenever a visual ray touches a visible object the object must somehow be seen. To accommodate the logical impossibility of the image location, Ptolemy modified the role of vision; the mind became an active agent, taking part in the phenomenon under study.[39] For Ptolemy the visual faculty, namely, the mind, transposes the image to the surface of the concave spherical mirror whenever the cathetus and the incident ray fail to intersect or the intersection occur behind the eye. Ptolemy therefore “transferred” the image to a location where, paradoxically, it is not (Smith, 1996, p. 194). In this methodology optics includes the activity of the mind; it is therefore essentially qualitative in spite of the geometrical analysis. Ptolemy's practice was based on sight-focused optical analysis of a particular case; this kind of optics mixes categories, the formal with the visual. Put succinctly, Ptolemy was very close, so to speak, to the phenomenon; he was true to optics as a visual science.

---

[36] Della Porta, 1589, Bk. 17. p. 271; Della Porta, 1593, Bk. 2, pp. 40–41; Smith, 2019, pp. 87–91. The many references Della Porta made in *Magiae natvralis* to his optical study (“in nostris opticis”) suggests that *De refractione* was available, perhaps in a manuscript form, earlier than this formal date of publication. On Della Porta's experiments with optical objects and his optical writings, see Borrelli, 2014, pp. 39–61.

[37] Della Porta, 1593, Bk. 2, pp. 36–41; Smith, 2019, pp. 79–91.

[38] For background discussion on Ptolemy's practical philosophy, see Neugebauer, 1969, pp. 191–206; Pedersen, 1974, pp. 26–32; Lindberg, 1976, pp. 1–13; Taub, 1994, pp. 19–37; Smith, 1996. pp. 21–30; Smith, 1999, pp. 23–49.

By contrast, Della Porta's study of reflection in concave spherical mirror was based on the assumption that there is a reciprocal relation between reflection in concave spherical mirror and refraction in glass sphere (Della Porta, 1593, Bk. 2, p. 35). This fundamental presupposition is categorically different from the one which Ptolemy assumed, namely, the cathetus principle. Della Porta explored experimentally the phenomenon of reflection in concave spherical mirror and refraction in glass sphere. He sought to account for what the optical element (e.g., concave spherical mirror and glass sphere) does rather than what the eye sees. As a result of his optical experiments with concave spherical mirrors in various setups, Della Porta discovered that the point of inversion is *not* located where the point of burning (i.e., the focal point) is. This discovery allowed Della Porta to exclude the point of inversion (visually dependent) from his discussion and to develop a novel analysis based solely on the physics of the setup. Visual perception was “out”; the physics of the optical element was rendered “in”. In other words, Della Porta eliminated visual considerations from his study of concave spherical mirrors and addressed only the physics of the setup. Unlike Ptolemy, Della Porta considered only geometrical-physical aspects of the optical arrangement. Della Porta thus introduced a new theory of concave spherical mirror. The link Della Porta established between geometrical analysis and systematic procedures for physically reproducing optical phenomena attest to his methodology—one needs to test in physical situations the relevant geometrical propositions. This scientific methodology was based on light-focused optical analysis directed at how light interacts physically with the reflecting mirror. It led Della Porta to discover what one could have called a law of nature, that is, the law of reflection in spherical concave mirror. Indeed, in *De refractione* Della Porta referred to this scheme as a law.[40] We are persuaded that to all intents and purposes Della Porta's formulation is nomological; he assumed optics to be based on universally binding rules stated in quantitative terms.[41] In sum, Della Porta was true to optics as a physical science.

[39] For Ptolemy, the physiological and psychological aspects of visual perception is “naturally” part and parcel of vision. It involves judgment about what is seen and this is based on a mental process of evaluation. This is the conceptual framework within which Ptolemy faced the problem. When mathematics failed to account for the place where the image should be located, psychology of visual perception came to rescue as an internal explanatory source available in scientific methodology.

Ptolemy and Della Porta shared the notion that geometry lies at the foundations of optics. However, while their respective observational and theoretical practices were similar in that they were geometrical, their scientific inquiries rested on different conceptions of optics. Ptolemy offered a procedure by which one could examine geometrically the location of the images he discussed in his optical analysis. He aimed at examining objects and phenomena as they actually appear under regularly defined conditions. Ptolemy considered the optical phenomenon he studied a single event explained by the cathetus principle. As he dealt with one case at a time, Ptolemy's account may be characterized as “particular”. He considered his experiments illustrations of common experience, knowledge claims related to particular cases and not arguments aimed at justifying universal propositions of optics.[42] Della Porta's study of reflection presupposed that reflection and refraction were essentially the same optical phenomenon devoid of visual considerations. Della Porta experimented on effects produced by concave spherical mirror and glass spheres under various regular conditions. He was concerned with the physical performance of the mirror and the glass sphere, and developed mathematical schemes with which he determined the physical properties of these optical elements. As he dealt with optical phenomena generated in a variety of optical setups, Della Porta's account may be characterized as “general”. His empirical approach was of a different nature from the common experience of Ptolemy. Della Porta’s geometrical analysis of his newly obtained observational data resulted in a universal proposition—independently of the visual faculty—about the optical phenomenon he studied.

What was conducive in Della Porta's scientific practice to a nomological approach to optics? Essentially, Della Porta’s theory of reflection makes the following two claims of proportionality: For a given incident ray, (1) the larger the radius of an optical interface is, the further away the converging/diverging ray will intersect the optical axis from the optical interface; and (2) the larger the radius of an optical interface is, the smaller the angle of reflection will be. Della Porta was critical of his predecessors who had not accounted accurately for the path of the solar ray as it is reflected and refracted in optical elements. These claims of proportionality, linking as they mathematically do the physical properties of the optical elements and the optical behavior of the ray, made it possible to think nomologically about these ubiquitous phenomena.

[40] Della Porta, 1593, pp. 3, 7, 84; Smith, 2019, pp. 8–9, 14–15, 172–173: *Refractionis enim lege*, *Refractionum principia*, and *Lex refractionis*, respectively.

[41] On the nomological view of nature, see Kedar, 2016, p. 71. Cf., Zik and Hon, 2012; 2017a.

[42] On the meaning of phenomenon and observation and the way the concept of common experience was evolved over time in astronomy, optics, and expertise, see Dear, 1995, pp. 11–31, 46–62.